\documentclass[12pt]{article}
\usepackage{amsmath}
\usepackage[pdftex]{graphicx}

\begin{document}

\section*{Merton Investment Problems in Finance and Insurance for the Hawkes-based Models\\}
{\bf Anatoliy Swishchuk\\}
{\it Department of Mathematics \& Statistics\\
University of Calgary\\
Calgary, Alberta, canada, T2N 1N4\\
E-mail: aswish@ucalgary.ca\\
}

{\bf Abstract:} We show how to solve Merton optimal investment stochastic control problem for  Hawkes-based models in finance and insurance, i.e., for a wealth portfolio $X(t)$ consisting of a bond and a stock price described by general compound Hawkes process (GCHP), and for a capital $R(t)$ of an insurance company with the amount of claims described by the risk model based on GCHP.  The novelty of the results consists of the new Hawkes-based models and in the new optimal investment results in finance and insurance for  those models.

{\bf Keywords:} Merton investment problem; optimal control; Hawkes process; general compound Hawkes process; LLN and FCLT; risk process; optimal investment in finance; optimal investment in insurance; diffusion approximation.

\section{Introduction}

Merton optimal investment and consumption stochastic problem is one of the most studied classical problem in finance (\cite{M1,M2,M3,B,KS}). In this paper, we will show how to solve the Merton optimal investment stochastic control problem for  Hawkes-based models in finance and insurance, i.e., for a wealth portfolio $X(t)$ consisting of a bond and a stock price described by general compound Hawkes process (GCHP) (\cite{S2,SH,S3}), and for a capital $R(t)$ of an insurance company with the amount of claims described by risk model based on GCHP (\cite{S,SZZ}). 

Namely, we will show how to solve the following two portfolio investment problems: 

1)  Merton portfolio optimization problem in finance  (\cite{M1,M2}) aims to find the optimal investment strategy for the investor with those two objects of investment, namely risk-less asset (e.g., a bond or bank account), paying a fixed rate of interest $r,$ and a number of risky assets (e.g., stocks) whose price GCHP. In this way, in our case, we suppose that $B_t$ and $S_t$ follows the following dynamics, respectively:
$$
\left\{
\begin{array}{rcl}
B_t&=&B_0\exp\{rt\}\\
S_t&=&S_0\exp\{G(t)\},\\
\end{array}
\right.
$$
\newpage
where $G(t):=\sum_{i=1}^{N(t)}a(X_i)$ is the GCHP, $X_i$ is a discrete-time Markov chain (MC) with finite or infinite states, $r>0$ is the interest rate, and $a(x)$ is a continuous and finite function on $X.$  We note, that the justification of using HP in finance may be found in \cite{DZ}, and using GCHP that based on HP $N(t)$ and $a(X_i)$ may be found in \cite{SH,HS}. The model for a stock price $S(t)$ that based on GCHP is a new and original in this paper.
The investor starts with an initial amount of money, say $X_0=x,$ and wishes to decide how much money to invest in risky and risk-less assets to maximize the final wealth $X_t$ at the maturity $T;$ 

2) Merton portfolio optimization problem in insurance aims to find an optimal investment for the capital  $R(t)$ of an insurance company at time $t$ ($R(t)$ is actually the risk model based on general compound Hawkes process (GCHP) (\cite{S,SZZ})), when an investor decides to invest some capital $A(t)$ in risky assets (e.g., stocks) and the rest, ($R(t)-A(t))$ in risk-free assets (e.g., bonds or bank account). We note, that the risk  model $R(t),$ based on GCHP, has the following representation:
$$
R(t)=u+ct-\sum_{i=1}^{N(t)}a(X_i),
$$
where $R(t)=u$ is the initial capital, $c>0$ is the premium rate, $N(t)$ is the Hawkes process, $X_i$ is a discrete-time finite or infinite state Markov chain with state space $X=\{1,2,3,...,N\}$ or $X=\{1,2,...,N,...\},$ respectively,  and $a(x)$ is a continuous and finite function on $X.$ We note, that the justification for the Hawkes-based risk model in the form of the above equation may be found in \cite{SZZ}.

The investor starts with an initial capital, say $R(0)=u,$ and wishes to decide how much money to invest in risky and risk-less assets to maximize the capital $R(t).$

The Merton optimal investment and consumption stochastic problem in finance was first considered in the seminal papers of Merton, \cite{M1,M2}. General description of the problem and coverage of most today's problems and methods may be found in \cite{FR, FS, B,M3,KS}. 

The first papers on stochastic optimal control in insurance appeared relatively recently, e.g., we would like to mention the papers written by Martin-L\"{o}f \cite{ML}, Brockett and Xia \cite{BX}, Browne \cite{B1}, to name a few. Since then many papers and books were written on this topic including \cite{AT, HT, Sch}. Financial control methods applied in insurance, such as e.g. in the control and the management  of the specific risk insurance companies, are described in \cite{Hipp}.  Risk theory for the compound Poisson process that is perturbed by diffusion was considered in \cite{DG}. An analogue of the Cramer-Lundberg approximation in the optimal investment case was studied in \cite{G2}. Asymptotic ruin probabilities and optimal investment were investigated in \cite{GGS}. Optimal risk distribution control model with application to insurance was studied in \cite{T}. Applications of stochastic processes in insurance and finance may be found in \cite{RSST}. Many mathematical methods and aspects in risk theory may be found in \cite{B2,G1,G}.

Hawkes process was first introduced in \cite{H1,H2}. Good introduction into Hawkes  processes and their properties may be found in \cite{L}. GCHP and regime-switching GCHP were first introduced in \cite{S3} and studied in details using real data in \cite{SRES,S2,HS,SH}. Risk model based on GCHP was first introduced in \cite{S1} and described in details in \cite{S}. Applications of the risk model based on GCHP to empirical data and optimal investment problem were considered in \cite{SZZ}.

The paper is organized as follows. Section 2 is devoted to the definitions and properties of Hawkes process and general compound Hawkes processes, and LLN (Law of Large Numbers) and FCLT (Functional Central Limit Theorem) for them. Section 3 deals with Merton investment problem in finance for the stock price described by GCHP, and Section 4 deals with Merton investment problem in insurance for the risk model based on GCHP. Section 5 concludes and describes the future work.

\section{General Compound Hawkes process}

\subsection{Hawkes Process}

{\bf Definition (One-dimensional Hawkes Process) (\cite{H1,H2})}. The one-dimensional Hawkes process is a point process $N(t)$ which is characterized by its intensity $\lambda(t)$ with respect to its natural filtration:
$$
\lambda (t)=\lambda+\int_{0}^t\mu(t-s)dN(s),
$$
where $\lambda>0,$ and the response function or self-exciting function $\mu(t)$ is a positive function and satisfies $\int_0^{+\infty}\mu(s)ds:=\hat\mu<1.$ 

If $(t_1,t_2,...,t_k)$ denotes the observed sequence of past arrival times of the point process up to time $t,$ the Hawkes conditional intensity is
$$
\lambda(t)=\lambda+\sum_{t_k<t}\mu(t-t_k).
$$

The Hawkes process is a self-exciting simple point process first introduced by A. Hawkes in 1971 (\cite{H1,H2}). The future evolution of a self-exciting point process is influenced by the timing of past events. 

The process is non-Markovian except for some very special cases (e.g., exponential self-exiting function $\mu(t)$). Thus, the Hawkes process depends on the entire past history and has a long memory. 

The Hawkes process has wide applications in neuroscience, seismology, genome analysis, finance, insurance, and many other fields. 

The constant $\lambda$ is called the background intensity and the function $\mu(t)$ is sometimes also called the excitation function. 

We suppose that $\mu(t)\not= 0$ to avoid the trivial case, which is, a homogeneous Poisson process. Thus, the Hawkes process is a non-Markovian extension of the Poisson process.

The interpretation of the above equation for $\lambda(t)$ is that the events occur according to an intensity with a background intensity $\lambda$ which increases by $\mu(0)$ at each new event then decays back to the background intensity value according to the function $\mu(t).$ 

Choosing $\mu(0)>0$ leads to a jolt in the intensity at each new event, and this feature is often called a self-exciting feature, in other words, because an arrival causes the conditional intensity function $\lambda(t)$ in (1)-(2) to increase then the process is said to be self-exciting.

The following LLN and CLT for HP may be found in \cite{BDHM}. The convergences are considered in weak sense for the Skorokhod topology.

{\bf LLN for HP} (\cite{BDHM}). Let $0<\hat\mu:=\int_0^{+\infty}\mu(s)ds<1.$ Then
$$
\frac{N(t)}{t}\to_{t\to+\infty}\frac{\lambda}{1-\hat\mu}.
$$
{\it Remark 1:} By LLN $N(t)\approx \frac{\lambda}{1-\hat\mu}t$ for large $t.$

{\bf FCLT for HP} (\cite{BDHM}).  Under LLN and $\int_0^{+\infty}s\mu(s)ds<+\infty$  conditions
$$
P\Big(\frac{N(t)-\lambda t/(1-\hat\mu)}{\sqrt{\lambda t/(1-\hat\mu)^3}}<y\Big)\to_{t\to+\infty}\Phi(y),
$$
where $\Phi(\cdot)$ is the c.d.f. of the standard normal distribution.

{\it Remark 2:} By FCLT $N(t)\approx \frac{\lambda}{1-\hat\mu}t+\sqrt{\lambda/(1-\hat\mu)^3)}W(t)$ for large $t,$ where $W(t)$ is a standard Wiener process (see \cite{BDHM}).

Remarks 1 and 2 above give the ideas about the averaged and diffusion approximated HP on a large time interval.
  
\subsection{General Compound Hawkes Process}

{\bf Definition (General Compound Hawkes Process).} General compound Hawkes Process is defined as (\cite{S2,SH,S3})
$$
S(t)=S(0)+\sum_{i=1}^{N(t)}a(X_i).
$$
Here, $X_i$ is a discrete-time finite or infinite state Markov chain with state space $X=\{1,2,...,N\},$ or $X=\{1,2,...,N,...\},$  respectively, $a(x)$ is a continuous and bounded function on $X,$ and  $N(t)$ is a Hawkes process with intensity $\lambda (t)>0,$ independent of $X_i.$

This general model is rich enough to:

$\bullet$ incorporate non-exponential distribution of inter-arrival times of orders in HFT or claims in insurance (hidden in $N(t)$)

$\bullet$ incorporate the dependence of orders or claims (via MC $X_i$)

$\bullet$ incorporate clustering of of orders in HFT or claims (properties of $N(t)$)

$\bullet$  incorporate order or claim price changes different from one single number (in $a(X_i)$).

This model is also very general to include:

{\bf -in finance}: 

$\bullet$ compound Poisson process:
$
S_t=S_0+\sum_{k=1}^{N(t)}X_k,
$  
where $N(t)$ is a Poisson process and $a(X_k)=X_k$ are i.i.d.r.v.

$\bullet$ compound Hawkes process (\cite{SRES}): 
$
S_t=S_0+\sum_{k=1}^{N(t)}X_k,
$  
where $N(t)$ is a Hawkes process and $a(X_k)=X_k$ are i.i.d.r.v.

$\bullet$ compound Markov renewal process: $
S_t=S_0+\sum_{k=1}^{N(t)}a(X_k),
$  
where $N(t)$ is a renewal process and $X_k$ is a Markov chain;

{\bf -in insurance}: 

$\bullet$ classical Cramer-Lundberg model: $a(X_i)=X_i,$ $X_i$ are i.i.d.r.v., and  $\mu(t)=0$ (then $N(t)$ is a poisson process);

$\bullet$ Sparre-Andersen model: $a(X_i)=X_i,$ $X_i$ are i.i.d.r.v., $\mu(t)=0,$ and $N(t)$ is a renewal process;

$\bullet$ Markov-modulated model: $a(X_i)=X_i,$ $X_i$ are i.i.d.r.v., $\lambda(t)=\lambda(X(t)),$ where $X(t)$ is a MC; we call this model regime-switching risk model based on GCHP (\cite{S2,S3}).


\subsection{LLN and FCLT for GCHP}

{\bf Lemma (LLN for GCHP) (\cite{S2,SH,S3})}. Let $\hat\mu:=\int_0^{+\infty}\mu(s)ds<1,$ and  Markov chain $X_i$ is ergodic with stationary probabilities $\pi^*_i.$ Then the GCHP $S_{nt}$ satisfies the following weak convergence in the Skorokhod topology:
$$
\frac{S(nt)}{n}\to_{n\to+\infty}a^*\cdot\frac{\lambda}{1-\hat\mu}t,
$$
or
$$
\frac{S(t)}{t}\to_{t\to+\infty}a^*\cdot\frac{\lambda}{1-\hat\mu}.
$$
Here: $ a^*$ is defined as
$a^*:=\sum_{i\in X}\pi^*_ia(i),$ where $\pi_i$ are ergodic probabilities for Markov chain $X_i.$

{\bf Theorem 1 (FCLT (or Jump-Diffusion Limit) for GCHP)  (\cite{S2,SH,S3}).} Let $X_k$ be an ergodic Markov chain with $n$ states $\{1,2,...,n\}$ and with ergodic probabilities $(\pi^*_1,\pi^*_2,...,\pi^*_n).$ Let also $S_t$ be LGCHP, and $
0<\hat\mu:=\int_0^{+\infty}\mu(s)ds<1\quad and\quad \int_0^{+\infty}\mu(s)sds<+\infty.$ Then
$$
\frac{S(nt)-N(nt)\cdot a^*}{\sqrt{n}}\to_{n\to+\infty}\sigma\sqrt{\lambda/(1-\hat\mu)}W(t),
$$
in weak sense for the Skorokhod topology, where $W(t)$ is a standard Wiener process, $\sigma$  is defined as:

$(\sigma)^2:=\sum_{i \in X} \pi^*_iv(i)$
$$
\begin{array}{rcl}
v(i)&=& b(i)^2+\sum_{j\in X}(g(j)-g(i))^2P(i,j)\\
&-&2b(i)\sum_{j\in X}(g(j)-g(i))P(i,j),\\
b&=&(b(1),b(2),...,b(n))',\\
b(i):&=&a(i)-a^*, \\
g:&=&(P+\Pi^*-I)^{-1}b,\\
\end{array}
$$
$P$ is a transition probability matrix for $X_k,$ i.e., $P(i,j)=P(X_{k+1}=j|X_k=i),$ 

$\Pi^*$ denotes the matrix of stationary distributions of $P,$ and $g(j)$ is the jth entry of $g.$

{\it Remark 3.} The formulas for $a^*$ and $\sigma$ look much simpler in the case of two-state Markov chain $X_i=\{-\delta,+\delta\}:$

$$
a^*:=\delta(2\pi^*-1)\quad and\quad  (\sigma^*)^2:=4\delta^2\Big(\frac{1-p'+\pi^*(p'-p)}{(p+p'-2)^2}-\pi^*(1-\pi^*)\Big),
$$
$(p,p')$ are transition probabilities of Markov chain $X_k,$ and $\pi^*_1=\pi^*,\quad \pi_2^*=1-\pi^*.$ 

From CLT for HP, sec. 3.1, and from Theorem 1 above follows the following FCLT for GCHP (pure jump diffusion limit).

{\bf Theorem 2 (FCLT (or Pure Diffusion Limit) for GCHP (\cite{S2,SH,S3,SZZ}) .} Let $X_k$ be an ergodic Markov chain with $n$ states $\{1,2,...,n\}$ and with ergodic probabilities $(\pi^*_1,\pi^*_2,...,\pi^*_n).$ Let also $S_t$ be LGCHP, and $
0<\hat\mu:=\int_0^{+\infty}\mu(s)ds<1\quad and\quad \int_0^{+\infty}\mu(s)sds<+\infty.$ Then
$$
\frac{S(t)-a^*\frac{\lambda}{1-\hat\mu}t}{\sqrt{t}}\to_{t\to+\infty}\bar\sigma N(0,1),
$$
in weak sense for the Skorokhod topology, where $N(0,1)$ is the standard normal c.d.f., and $\bar\sigma$ is defined as:
$$
\bar\sigma=\sqrt{(\sigma^*)^2+\Big(a^*\sqrt{\frac{\lambda}{(1-\hat\mu)^3}}\Big)^2},
\eqno{(Vol)}
$$
where $\sigma^*=\sigma\sqrt{\lambda/(1-\hat\mu)},$ $\sigma$ and $a^*$ are defined in Theorem 1 and Lemma above, respectively.

{\it Remark 4.} The Theorem 2 implies that $S(t)$ can be approximated by the pure diffusion process:
$$
S(t)\approx S(0)+a^*\frac{\lambda}{1-\hat\mu}t+\bar\sigma W(t),
$$
where $W(t)$ is a standard Wiener process. This Remark 4 gives the idea about the pure diffusion approximation of GCHP on a large time interval.

{\it Remark 5.} We note, that the rate of convergence in the Theorem 2 is $C(T)/\sqrt{t},$ $0\leq t\leq T,$ where $C(T)>0$ is a constant (\cite{SZZ}). Thus, the error of approximation for $S(t)$ in Remark 4 is small for large $t.$

\section{Merton Investment Problem in Finance for the Hawkes-based Model}

Let us consider Merton portfolio optimization problem. We suppose that $B_t$ and $S_t$ follows the following dynamics, respectively:
$$
\left\{
\begin{array}{rcl}
B_t&=&B_0\exp\{rt\}\\
S_t&=&S_0\exp\{G(t)\},\\
\end{array}
\right.
\eqno{(1)}
$$
where $G(t):=\sum_{i=1}^{N(t)}a(X_i)$ is the GCHP, $r>0$ is the interest rate.

We note, that the justification of using HP in finance may be found in \cite{DZ}, and using GCHP that based on HP $N(t)$ and $a(X_i)$ may be found in \cite{SH,HS}. The model for a stock price $S(t)$ that based on GCHP is a new and original in this paper.

The investor starts with an initial amount of money, say $X_0=x,$ and wishes to decide how much money to invest in risky and risk-less assets to maximize the expected utility of the terminal wealth $X_t$ at the maturity $T,$ i.e., $X_T.$ 



We denote by $n(t):=(n_B(t),n_S(t))$ an investor portfolio, where $n_B(t)$ and $n_S(t)$ are the amounts in cash invested in the bonds and the risky assets, respectively. The value $X(t)$ at time $t$ of such portfolio  is 
$$
X(t)=n_B(t)+n_S(t).
$$
We suppose that our portfolio is admissible, i.e., $X(t)\geq 0,$ a.s., $0\leq t\leq T,$ and self-financing, i.e.,
$$
dX(t)=n_B(t)\frac{dB(t)}{B(t)}+n_S(t)\frac{dS(t)}{S(t)}.
$$  

 Suppose that $G(t):=\sum_{i=1}^{N(t)}a(X_i)$ follows FCLT when $t\to+\infty,$ (\cite{S,SZZ}), thus $G(t)$ can be approximated as (see sec. 3.3)
 $$
 G(t)\approx a^*\frac{\lambda}{1-\hat\mu}t+\bar\sigma W(t),
 \eqno{(2)}
  $$
where $\hat\mu:=\int_0^{+\infty}\mu(s)ds<1,$ $a^*$ is a average of $a(x)$ over stationary distribution of MC $X_i,$ $\lambda$ is a background intensity, $\bar\sigma>0$ is defined in sec. 3.3. For exponential decaying intensity $\hat\mu=\alpha/\beta.$

Thus, $S(t)$ in (1) can be presented in the following way:
$$
S(t)=S(0)e^{a^*\frac{\lambda}{1-\hat\mu}t+\bar\sigma W(t)}.
\eqno{(3)}
$$
Using $It\hat o$ formula we can get:
$$
dS(t)=S(t)[(a^*\frac{\lambda}{1-\hat\mu}+\frac{\bar\sigma^2}{2})dt+\bar\sigma dW(t)].
\eqno{(4)}
$$
Then the  change of the wealth process $X_t$ can be rewritten in the following way, taking into account (1)-(4):
$$
\begin{array}{rcl}
dX_t&=&n_B(t)\frac{dB(t)}{B(t)}+n_S(t)\frac{dS(t)}{S(t)}\\
&=&n_B(t)rdt+n_S(t)[(a^*\frac{\lambda}{1-\hat\mu}+\frac{\bar\sigma^2}{2})dt+\bar\sigma dW(t)]\\
&=&rX(t)dt+n_S(t)(a^*\frac{\lambda}{1-\hat\mu}+\frac{\bar\sigma^2}{2}-r)dt+n_S(t)\bar\sigma dW(t).\\
\end{array}
\eqno{(5)}
$$

Let $\pi(t):=n_S(t)/X(t)$ be the portion of wealth invested in the assets/stocks at time $t.$

 Then, from (1)-(5), we have the following expression for $dX(t):$
 $$
 \begin{array}{rcl}
dX_t&=&rX(t)dt+n_S(t)(a^*\frac{\lambda}{1-\hat\mu}+\frac{\bar\sigma^2}{2}-r)dt+n_S(t)\bar\sigma dW(t).\\
&=&dX(t)=X(t)[(r+\pi_t(a^*\frac{\lambda}{1-\hat\mu}+\frac{\bar\sigma^2}{2}-r))dt+\pi_t\bar\sigma dW(t)))]\\
\end{array}
\eqno{(6)} 
$$
Finally, after replacing $X(t)$ with $X^{\pi}(t),$ to stress the dependence of $X(t)$ on $\pi_t,$ from (6) we have the following equation for $dX^{\pi}(t):$
$$
dX^{\pi}(t)=X^{\pi}(t)[(r+\pi(t)(a^*\frac{\lambda}{1-\hat\mu}+\frac{\bar\sigma^2}{2}-r))dt+\pi(t)\bar\sigma dW(t)))].
\eqno{(7)}
$$
Our main goal is to solve the following optimization problem:
$$
\max_{\pi}E[U(X_T^{\pi})|X_0=x],
$$
meaning to maximize  the wealth/value function or performance criterion \\$E[U(X_T^{\pi})|X_0=x],$ where $U(x)$ is a utility function. 

To find optimal $\pi,$ we follow the standard procedure in this case (see \cite{B,KS}). For the utility function we take the logarithmic one, $U(x)=\log(x).$  Therefore, we have to maximize $\max_{\pi}E[\log(X_T^{\pi})|X_0=x],.$ Solving the equation (7) and maximizing non-martingale term in the exponent for the solution, we can find the optimal investment solution $\pi^*(t):$
$$
\pi^*(t)=\frac{a^*\frac{\lambda}{1-\hat\mu}+\frac{\bar\sigma^2}{2}-r}{\bar\sigma^2},
$$
where
$$
\bar\sigma=\sqrt{(\sigma^*)^2+\Big(a^*\sqrt{\frac{\lambda}{(1-\hat\mu)^3}}\Big)^2},
$$
and $\sigma^*$ and $a^*$ are defined in sec. 3.3.

{\it Remark 6.} As we can see from the expression for $\pi^*(t),$ the optimal investment solution depends on all parameters of the Hawkes-based model, namely, Hawkes process's parameters $\lambda$ and $\mu(t),$ Markov chain $X_i$ and function $a(x)$ through $a^*.$  

\section{Merton Investment Problem in Insurance for the Hawkes-based Risk Model}

Let us consider $R(t)$ as the risk model based on GCHP, namely, 
 $$
R(t)=u+ct-\sum_{i=1}^{N(t)}a(X_i).
$$
Here, $X_i$ (claim sizes) is a finite state Markov chain, $a(x)$ is a continuous and bounded function on $X=\{1,2,...,N\}$- space state for $X_i,$ and $N(t)$ is a Hawkes process with intensity $\lambda (t)>0,$ independent of $X_i,$ and satisfying:
$$
\lambda(t)=\lambda+\int_0^t\mu(t-s)dN(t).
$$
Here, $\mu(t)$ is self-exiting function.

We note, that the justification for the Hawkes=based risk model in the form of the above equation may be found in \cite{SZZ}.

Since we will consider optimization for a first insurer, thus we will concentrate on problems with infinite planning horizon. 

Let $A(t)$ be an amount invested in a risky asset, and suppose that the price $S(t)$ of the risky asset follows GBM, i.e.,  

$$dS(t)=S(t)(adt+bdW(t)),$$
where $a$ is a real constant, $b>0.$ 

Further, the leftover, $R(t)-A(t)>0,$ is invested in  a bank account (or bonds) with interest rate $r>0,$  thus 
$$d(R(t)-A(t))=r(R(t)-A(t))dt.$$ 

 {\bf Merton Investment Problem in Insurance}
 
 Let $\beta(t):=A(t)/S(t)$ be the number of assets held at time $t.$ Then the position of the insurer has the following dynamics:
 $$
 dR(t)=rR(t)dt+\beta(t)dS(t)-r\beta(t)S(t)dt+cdt-d(G(t)).
 $$

Therefore, the  dynamics for $R(t)$ is (taking into account all above equations for $S(t),$ $d(R(t)_A(t))$ and dR(t)): 
$$
\begin{array}{rcl}
dR(t)=rR(t)dt+A(t)(adt+bdW(t))-rA(t)dt+cdt-d(G(t))\\
\end{array}
$$
Here, $G(t)=\sum_{i=1}^{N(t)}a(X_i).$ 

Let $\pi(t):=A(t)/R(t)$ be the fraction of the total wealth $R(t)$ invested in the risky assets. 

Then, we can rewrite the equation for $dR(t)$ in the following way (we use notation $R^{\pi}(t)$ to stress dependence of $R(t)$ from $\pi$):
 $$
 \begin{array}{rcl}
 dR^{\pi}(t)&=&rR^{\pi}(t)dt+A(t)(adt+bdW(t))-rA(t)dt+cdt-d(G(t))\\ 
 &=&rR^{\pi}(t)dt+A(t)((a-r)dt+bdW(t))+cdt-d(G(t))\\
 &=&R^{\pi}(t)[(r+\pi(t)(a-r))dt+\pi(t)bdW(t)]+cdt-d(G(t)),\\
 \end{array}
$$
 where $W(t)$ is a standard Brownian motion. 
 
 As for the control at time $t$ we will take the function $\pi(t),$ i.e., the fraction of the total wealth $R^{\pi}(t)$ which should be invested in risky assets. 
 
 We will show how to find the optimal strategy $\pi(t),$ which maximizes our expected utility function, $E[U(R^{\pi}(t))|R^{\pi}(0)=r],$ where $U(r)$ is a special utility function. 
 
 We suppose that $G(t)=\sum_{i=1}^{N(t)}a(X_i)$ follows FCLT when $t\to+\infty,$ (\cite{S,SZZ}), thus $G(t)$ can be approximated as (see sec. 3.3)
 $$
 G(t)\approx a^*\frac{\lambda}{1-\hat\mu}t+\bar\sigma W_1(t),
  $$
where $\hat\mu:=\int_0^{+\infty}\mu(s)ds<1,$ $a^*$ is a average of $a(x)$ over stationary distribution of MC $X_i,$ $\lambda$ is a background intensity, $\bar\sigma$ is defined in sec. 3.3. For exponential decaying intensity $\hat\mu=\alpha/\beta.$
 
 Here, $W_1(t)$ is a Wiener process independent of $W(t)$ (the case for correlated $W_1(t)$ with $W(t)$ i.e., such that $[W(t),W_1(t)]=\rho dt,$ can be considered as well with some modifications).
 
 We suppose that $c>a^*\frac{\lambda}{1-\hat\mu}$ (safety loading condition).
  
 After substituting (2) into (1) for dR(t) we get:
 $$
 \begin{array}{rcl}
 dR^{\pi}(t)&=&[R^{\pi}(t)(r+(a-r)\pi(t))+(c-a^*\frac{\lambda}{1-\hat\mu})]dt\\
 &+&\sqrt{R^2(t)b^2\pi^2(t)+\bar\sigma^2}dW_2(t), \\
 \end{array} 
 $$
 where $W_2(t)$ is a standard Wiener process independent of $W(t)$ and $W_1(t).$
 
  Generator for $R^{\pi}(t)$ is (here, $R^{\pi}(0)=x$)
 $$
 \begin{array}{rcl}
 A^{\pi}&=&[x(r+(a-r)\pi(t))+(c-a^*\frac{\lambda}{1-\hat\mu})]\frac{\partial}{\partial x} \\
 &+&[(x^2b^2\pi^2(t)+\bar\sigma^2)/2]\frac{\partial}{\partial x^2} \\
 \end{array} 
 $$
 Thus, we have to  maximize $E_r[U(R^{\pi}(t))],$ where $U(r)$ is a utility function.
 
The HJB equation has the following form:
$$
\frac{\partial}{\partial r}v(t,r)+\sup_{\pi}[A^{\pi}v(t,r)]=0,
\eqno{(HJB)}
$$

where $v(t,r)=\sup_{\pi}E_r[U(R^{\pi}(t))].$

We take the  exponential utility function  $U(r)=-e^{-pr}, \quad p>0.$

We suppose that  $0<p<\frac{2xr}{\sigma^2}=\frac{2R^{\pi}(0)r}{\bar\sigma^2}.$

Solving the HJB equation we get the  optimal control $\pi(t):$
 
  $$
 \pi(t)=\frac{(a-r)}{xpb^2}
 $$
 
 where $x=R^{\pi}(0)>0,$ and $\pi(t)$ depends on $p.$  After finding 
 $$
 p=\frac{\theta+\sqrt{\theta^2+\bar\sigma^2(a-r)^2/b^2}}{\bar\sigma^2},
 $$ 
 where $\theta:=xr+(c-a^*\frac{\lambda}{1-\hat\mu}),$
 we can finally find that 
 $$
 \pi(t)=\frac{\bar\sigma^2(a-r)}{xb(\theta b+\sqrt{\theta^2b^2+\bar\sigma^2(a-r)^2})},
 $$
 where
$$
\bar\sigma=\sqrt{(\sigma^*)^2+\Big(a^*\sqrt{\frac{\lambda}{(1-\hat\mu)^3}}\Big)^2},
$$ 
$\sigma^*=\sigma\sqrt{\lambda/(1-\hat\mu)},$ $\sigma$ and $a^*$ are defined in Theorem 1 and Lemma, sec. 2.3, respectively.

 As we can see, the optimal control $\pi(t)=\pi$ does not depend on $t,$ thus is a constant, and contains all initial parameters of the risk model based on GCHP.
 
 {\it Remark 7.} As we can see from the expression for $\pi(t),$ the optimal control depends not only from interest rate $r,$ but also from all parameters of the Hawkes-based model, namely, Hawkes process's parameters $\lambda$ and $\mu(t),$ Markov chain $X_i$ and function $a(x)$ through $a^*,$ and the asset's parameters $a$ and $b.$   
 
 {\bf  Merton Investment Problem for Poisson-based Risk Model in Insurance}.
 
 {\it Corollary}: The  optimal control $\pi_P(t)$ for Poisson Risk Model is:
 
  $$
  \begin{array}{rcl}
 \pi_P(t)= \frac{\lambda E[X_i]^2(a-r)}{xb(b(xr+(c-\lambda EX_i))+\sqrt{b^2(xr+(c-\lambda EX_i))+\lambda E[X_i]^2(a-r)^2)})}.
\end{array}
 $$
Here: $\bar\sigma=\sqrt{\lambda E[X_i]^2}.$ 
 
\section{Discussion}

The future work will be devoted to numerical example based on real data, simulations and considering the case 
{\it  without diffusion approximation} for $R(t),$ by creating HJB equation for initial risk model $R(t)=R(0)+ct-\sum_{i=1}^{N(t)}a(X_i)$ and by solving this HJB equation for initial risk model base on GCHP. Probably we cannot avoid here numerical methods, because the HJB equation in this case cannot be solved exactly with a close form solution.\\

\section{Conclusions}

We described in this paper how to solve Merton optimal investment stochastic control problem for  Hawkes-based models in finance and insurance, i.e., for a wealth portfolio $X(t)$ consisting of a bond and a stock price described by general compound Hawkes process (GCHP), and for a capital $R(t)$ of an insurance company with the amount of claims described by the risk model based on GCHP. The novelty of the results consists of the new Hawkes-based models and in the new optimal investment results in finance and insurance for  those models.


\vspace{6pt} 




{\bf Acknowledgments}: The author thanks to NSERC for continuing support.






\begin{thebibliography}{999}
\bibitem{AT}Asmussen, S. and Taksar, M. (1997). Controlled diffusion models for optimal dividend payout. {\it Insurance: Mathematics and Economics}, 20, 1-15.

\bibitem{BDHM}Bacry, E.,  Delattre, S., Hoffmann, M. and Muzy, J. (2013): Some limit theorems for Hawkes processes and application to financial statistics. {\it Stochastic Processes and their Applications}, v. 123, issue 7, 2475-2499.
 
\bibitem{BX}Brockett, P., and Xia, X. (1995). Operations research in insurance: a review. {\it Trans. Act. Soc.}, XLVII, 7-80. 

\bibitem{B1}Browne, S. (1995.) Optimal investment policies for a firm with a random risk process: exponential utility and minimizing the probability of ruin. {\it Math. Operations Res.} 20, 937-958. 
 
 \bibitem{B2} B\"{u}hlmann, H. (1970). {\it Mathematical Methods in Risk Theory}. Springer, NY.
 
\bibitem{B}Bjork, T. (2009). {\it Arbitrage Theory in Continuous Time}. Oxford University Press. 3d ed.

\bibitem{DZ} DaFonseca, J. and Zaatour, R. (2013). Hawkes Process: Fast Calibration, Application to Trade Clustering, and Diffusive Limit. {\it J. Futures Markets}, v. 3, issue 6.

\bibitem{DG} Dufresne, F. and Gerber, H.U. (1991). Risk theory for the compound Poisson
process that is perturbed by diffusion. {\it Insurance: Math. Econom.}, 10, 51–59.

\bibitem{FR}Fleming, W.H. and Rishel, R.W. (1975). {\it Deterministic and stochastic optimal 
control}. Springer, New York. 
 
\bibitem{FS}Fleming, W.H. and Soner, M. (1993). {\it Controlled Markov processes and viscosity 
solutions}. Springer, New York. 
 
 \bibitem{GGS}Gaier, J., Grandits, P. and Schachermeyer, W. (2003). Asymptotic ruin probabilities
and optimal investment. {\it Annals of Applied Probability}, v.13, 3, 1054-1076.


\bibitem{G1}Gerber, H. U. (1979).  {\it An Introduction to Mathematical Risk Theory}. S.S. Huebner
Foundation Monographs, University of Pennsilvania.

\bibitem{G2}Grandits, P. (2003). An analogue of the Cramer-Lundberg approximation in the
optimal investment case. {\it Preprint}, Technical University Vienna.
 
 
 \bibitem{G}Grandell, J.  (1990). {\it Aspects of Risk Theory}. Springer-Verlag, NY.
 
\bibitem{H1}Hawkes, A. (1971). Spectra of some self-exciting and mutually-exciting point processes. {\it Biometrika}, 1971, 58, 83-90.

\bibitem{H2}Hawkes, A. (1971). Point spectra of some mutually-exciting point processes.
J. R. Stat. Soc. B, 1971b, 33, 438-443.

\bibitem{Hipp} Hipp, C. (2004). Stochasric optimal control in insurance. In: {\it Stochastic Methods in Finance}, Lecture Notes in Math., 1856, Springer, Ed.: Frittelli, M. and Runggaldier, W., 127-164.

\bibitem{HT}H$\oslash$jgaard, B. and Taksar, M. (1999). Controlling risk exposure and div-
idends pay-out schemes: Insurance company example. {\it Math.Finance},
Vol. 2, 153-182.

\bibitem{KS}Karatzas, I. and Shreve, S. (1998). {\it Methods of Mathematical Finance}. Springer, Probability Theory and Mathematical Modelling, v. 39.

\bibitem{L}Laub, P., Taimre, T. and Pollet, P. (2015). Hawkes Processes. {\it arXiv: 1507.02822v1}. 10 July 2015.
  
\bibitem{ML}Martin-L\"{o}f, A. (1994) Lectures on the use of control theory in insurance. {\it Scand. 
Actuarial J.}, 1-25. 
  
\bibitem{M1}Merton, R. (1969). Lifetime portfolio selection under uncertainty: The continuous-time case. {\it The Review of Econ. Stat.,} 247-257.
  
\bibitem{M2}Merton, R. (1971). Optimum consumption and portfolio rules in a continuous-time model. 
{\it J. of Economic Theory}. 3, 373-413.

\bibitem{M3}Merton, R. (1990). {\it Continuous-Time Finance.} Basil Blackwell Inc., Cambridge, MA.

\bibitem{RSST}Rolski, T. Schmidli, H. Schmidt, V. Teugels, J. (1998). {\it Stochastic Processes for
Insurance and Finance}. Wiley Series in Probability and Statistics

\bibitem{Sch}Schmidli, H. (2008). {\it Stochastic Control in Insurance}. Springer.

\bibitem{S1}Swishchuk, A. (2017). Risk processes based on general compound Hawkes processes. IME 2017 Congress. Abstracts, Vienna, Austria.

\bibitem{S3} Swishchuk, A. (2017). General compound Hawkes processes in LOB. {\it arXiv:1706.07459v2 [q-fin.MF]}. 2017. 

\bibitem{SHCS}Swishchuk, A., Cera, K., Hofmeister, T. and Schmidt, J.  (2017). General semi-
Markov model for limit order books. {\it Inern. J. Theoret. Applied Finance}, v. 20, 20
pages.

\bibitem{S}Swishchuk, A. (2018). Risk model based on general compound Hawkes process. {\it Wilmott}, v. 2018, Issue 94.

\bibitem{SRES}Swishchuk, A., Remillard, B. Elliott, R. and Chavez-Casillas, J. (2019). Compound Hawkes processes in limit order books. {\it Financial Mathematics, Volatility and Covariance Modelling.}  Routledge: Taylor and Francis Group. v. 2, 1st ed., 2019. (Editors: Chevallier, J., Goutte, S., Guerreiro, D., Saglio, S. and Sanhaji, B.)

 \bibitem{HS} Swishchuk, A. and He, Q. (2019). Quantitative and Comparative Analyses of Limit Order Books with General Compound Hawkes Processes. {\it Risks}, 2019, 7(4), 110.
 
\bibitem{S2}Swishchuk, A. (2020). Modelling of limit order books by general compound Hawkes processes with implementations. {\it Meth. Comput. Appl. Probab.}, 2020, 23, 299-428 (2021). 

\bibitem{SH}Swishchuk, A. and Huffman, A. (2020). General compound Hawkes processes in limit order books. {\it Risks}, 2020, 8(1), 28.

\bibitem{SZZ}Swishchuk, A., Zagst, R. and Zeller, G. (2020). Hawkes processes in insurance: Risk model, application to empirical data and optimal investment. {\it Insurance: Mathematics and Economics}, https://doi.org/10.1016/j.insmatheco.2020.12.005.

\bibitem{T}Taksar, M. (2000). Optimal Risk/Dividend Distribution Control Models: Applications
to Insurance. {\it Mathematical Methods of Operations Research}, 1, 1-42.


\end{thebibliography}


\end{document}